\documentclass[aps,prx,showpacs,twocolumn,reprint,superscriptaddress,nofootinbib]{revtex4-2}

\usepackage[dvips]{graphicx}
\usepackage{amsmath,amssymb,amsthm,mathrsfs,amsfonts,dsfont,mathtools,physics,adjustbox}
\usepackage{epsfig}
\usepackage{braket}

\usepackage[dvipsnames]{xcolor}

\usepackage{placeins}
\usepackage{bm}
\usepackage{enumerate}
\usepackage{color}
\usepackage{graphicx}
\usepackage{textcomp}
\usepackage{float}
\usepackage[version=4]{mhchem}
\usepackage{yquant, quantikz}
\useyquantlanguage{groups}
\usetikzlibrary{calc}
\usepackage{pgfplots}
\pgfplotsset{compat=newest}

\usepackage{enumitem}

\usepackage[normalem]{ulem}
\usepackage{comment}
\usepackage{xcolor}
\usepackage{silence}
\usepackage{amsthm}

\PassOptionsToPackage{breaklinks}{hyperref}
\usepackage{xurl}
\usepackage{hyperref}
\hypersetup{
    colorlinks,
    linkcolor={blue!50!black},
    citecolor={blue!50!black},
    urlcolor={blue!80!black}
}
\usepackage[capitalize]{cleveref}

\Crefname{theorem}{Theorem}{Theorems}
\theoremstyle{remark}

\begin{document}

\title{Space and Time Cost of Continuous Rotations in Surface Codes}

\newcommand{\qmaddress}{\affiliation{Quantum Motion, 9 Sterling Way, London N7 9HJ, United Kingdom}}
\newcommand{\oxddress}{\affiliation{Department of Materials, University of Oxford, Parks Road, Oxford OX1 3PH, United Kingdom}}
\newcommand{\mathinst}{\affiliation{Mathematical Institute, University of Oxford, Woodstock Road, Oxford OX2 6GG, United Kingdom}}

\author{Zhu Sun}
\email[zhu.sun@exeter.ox.ac.uk]{}
\qmaddress
\oxddress
\mathinst

\author{B\'alint Koczor}
\qmaddress
\mathinst

\begin{abstract}
While Clifford operations are relatively easy to implement in fault-tolerant quantum computers, continuous rotation gates remain a significant bottleneck in typical quantum algorithms. In this work, we ask the question: ``What is the most efficient approach for implementing continuous rotations in a surface code architecture?" Several techniques have been developed to reduce the T-count or T-depth of rotations, such as Hamming weight phasing and catalyst towers. However, these methods often require additional a number of ancilla qubits, and thus the ultimate cost function one needs to optimise against should rather be the total runtime or the total space required for performing a rotation. We explicitly construct surface code layouts for catalyst towers in two practical application examples in the context of option pricing: (a) implementing a phase oracle circuit, which is a ubiquitous subroutine in many quantum algorithms, and (b) state preparation using a variational quantum circuit. Our analysis shows that, at small and medium code distances, catalyst towers not only reduce the runtime but can also decrease the total spacetime volume of rotations. However, at large code distances, conventional Clifford$+$T synthesis may prove more efficient. Additionally, we note that our conclusions are sensitive to specific application scenarios and the choices of various parameters. Nevertheless, catalyst towers may be particularly advantageous for early fault-tolerant quantum applications, where low and medium code distances are assumed and a spacetime tradeoff is needed to reduce the runtime of individual circuit runs, such as in scenarios involving high circuit repetition counts.

\end{abstract}

\maketitle

\section{Introduction}
Quantum algorithms promise to solve a range of practically important problems significantly faster than classical computers. However, most typical quantum algorithms, from quantum signal processing type techniques to product formulas in quantum simulation, require a very large number of  continuous-angle rotations. Implementing these rotations usually represents a bottleneck as they need to be implemented as potentially deep sequences of Clifford and T gates in fault tolerance~\cite{zimboras2025myths}. A broad range of solutions have been developed in the literature to improve their efficiency from optimised gate synthesis protocols~\cite{ross2014optimal,bocharov2015efficient,kliuchnikov2023shorter} to minimising the cost of the simultaneous application of many rotations~\cite{sun2024, kivlichan2020improved}, while quasi-probability decompositions can be exploited when the aim is to estimate expected values~\cite{PRXQuantum.5.040352,PhysRevLett.132.130602}.

A particularly efficient method for achieving early quantum advantage leverages Hamming-weight phasing combined with phase gradients to implement continuous rotations at a relatively low T-gate cost~\cite{low2025fast}. However, this approach has two main limitations: it is restricted to parallel rotations with identical angles, and it enforces a ``serial execution" of these rotations, which can lead to increased algorithmic runtime when the number of parallel rotations is large.

Therefore, in this work, we focus on an alternative approach that enables a spacetime tradeoff: if more logical qubits are available, continuous rotations can be executed in parallel, leading to reduced runtime. Specifically, we consider catalyst circuits \cite{gidney2019efficient}, which can prepare resource states with continuous rotation angles while effectively consuming only four T gates. In addition, the approach requires access to identical, continuously rotated resource states, which are recovered at the end of the procedure -- hence the term ``catalyst". This technique was generalized in refs.~\cite{sun2024,wang2023option} by connecting multiple catalysts into a so-called catalyst tower, enabling the synthesis of a family of resource states. These states can be buffered and teleported using a repeat-until-success strategy.

The efficacy of the catalyst tower approach was demonstrated in a) oracle circuits whereby the aim is to perform the action on a computational state as $\ket{x}\mapsto\exp(i f(x))\ket{x}$)~\cite{sun2024} and b) time evolution circuits whereby we perform the action $\exp(-i \mathcal{H} t) |\psi\rangle$ on an arbitrary input state~\cite{kiumi2024te}. Indeed, catalysts in these applications have achieved significantly reduced T-gate counts and total circuit depths compared to conventional rotation-gate synthesis techniques. However, they also require additional ancillary qubits, and it remains an open question whether their advantages persist once all overheads associated with error correction and hardware constraints are taken into account. Furthermore, the cost of magic state distillation has been significantly reduced following recent breakthroughs \cite{litinski2019magic,gidney2019flexible,gidney2024magic}, which lowers the cost of T gates to a level comparable to that of fault-tolerant CNOT gates in certain regimes. For this reason, in the present work, we carry out a detailed resource analysis by constructing an explicit cost model that incorporates physical parameters such as the physical error rate and the number of physical qubits required per logical qubit.

Specifically, we assume a surface code architecture, which is the leading approach for building error-corrected quantum computers due to its practicality and compatibility with nearest-neighbour qubit connectivity—typical in most solid-state platforms. We explicitly lay out catalyst towers within a surface code architecture and compare their resource requirements to those of conventional gate synthesis methods in terms of physical qubit count and spacetime volume. We consider two practical application examples in the context of derivative pricing \cite{chakrabarti2021threshold,stamatopoulos2024derivative,sun2024}. First, we examine phase oracle circuits, which are ubiquitous in many quantum algorithms, and specifically focus on a piecewise linear approximation that is amenable to parallelization. Second, we consider a variational quantum circuit used to prepare multiple copies of Gaussian states, which is representative of a broader class of practically relevant variational circuits. While we find that catalyst towers can indeed reduce the total spacetime cost of rotations at low and intermediate code distances, we emphasize that the primary focus of the present application examples is the minimization of circuit depth, achieved through parallelization and by exploiting spacetime tradeoffs.

The remainder of this manuscript is organized as follows. In \cref{sec:background}, we review the necessary background on fault-tolerant architectures. Our layout and cost analysis of a phase oracle circuit are presented in \cref{sec:poc}. In \cref{sec:gaussian}, we describe the use of catalyst towers for applying rotations in variational circuits. We discuss our observations in \cref{sec:discussions} and conclude in \cref{sec:conclusion}.

\section{Background} \label{sec:background}

\subsection{Surface code}

Surface codes~\cite{bravyi1998quantum,kitaev2003fault} are the most promising candidate for error correction in solid-state platforms
given their robustness, practicality and logical operations that can be performed using local operations, such as in case of lattice surgery \cite{horsman2012surface}. 
Here we summarise some basic properties of surface codes while a detailed overview can be found in, e.g. ref.~\cite{fowler2012surface}.

In the surface code, physical qubits are arranged in a 2D square lattice,
thus constructing a code with code distance $d$ approximately requires $2d^2$ physical qubits to form a logical qubit patch.
A natural unit for clock time is the code cycle time, i.e. the time it takes to perform a full round of stabilizer checks while $d$ code cycles is often referred to as 1 time step. The logical error rate per logical qubit per code cycle can be approximated~\cite{fowler2018low} 
in terms of the physical error rate $p$  as
\begin{equation} \label{eqn:errorrate}
p_L(d)=0.1(100p)^{(d+1)/2}   .
\end{equation}

We assume a Clifford+T gate set, adopt Litinski's lattice surgery scheme in ref~\cite{litinski2019game},
and assume T-state distillation protocols as in ref.~\cite{litinski2019magic}. In particular, these assumptions enable the efficient movement of logical qubit patches in $d$ code cycles. However, as our primary aim in this work is to minimize the depth of the computation by simultaneously applying multiple operations onto the logical qubits,
we do not use the Pauli-based computation approach in ref.~\cite{litinski2019game}, which requires a sequential
application of magic resources to the logical qubits.
We also use the method in ref.~\cite{litinski2018lattice} to perform long-range CNOT operations and use the AutoCCZ factories from ref.~\cite{gidney2019flexible}.

\subsection{Catalyst towers} \label{sec:cata}

\begin{figure*}[t]
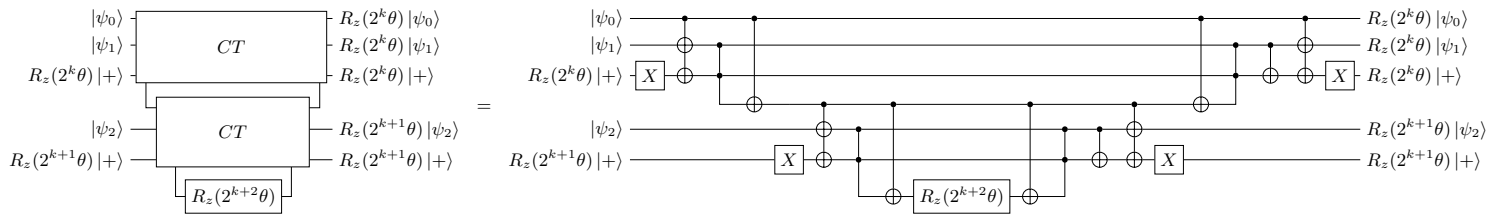

\hspace*{-1.1cm} 
\begin{adjustbox}{scale=0.77}
\begin{yquantgroup}
\registers{
    qubit {} a;
    qubit {} b;
    qubit {} c;
    qubit {} d;
    qubit {} e;
    qubit {} f;
    qubit {} g;}

\circuit{
    init {$\ket{\psi_0}$} a;
    init {$\ket{\psi_1}$} b;
    init {$R_z(2^k\theta)\ket{+}$} c;
    init {$\ket{\psi_2}$} e; 
    init {$R_z(2^{k+1}\theta)\ket{+}$} f;
    discard d,g;
    
    box {$\qquad\qquad CT \qquad\qquad$} (a,b,c);
    hspace {3.5mm} e;
    [name=ct1]
    box {$\qquad\quad CT \qquad\quad$} (d,e,f);
    hspace {8.5mm} g;
    [name=ct2]
    box {$R_z(2^{k+2}\theta)$} g;

    output {$R_z(2^k\theta)\ket{\psi_0}$} a;
    output {$R_z(2^k\theta)\ket{\psi_1}$} b;
    output {$R_z(2^k\theta)\ket{+}$} c;
    output {$R_z(2^{k+1}\theta)\ket{\psi_2}$} e;
    output {$R_z(2^{k+1}\theta)\ket{+}$} f;

    \draw (ct1)+(-1.5,0.85) |- + (-1.32,0.42);
    \draw (ct1)+(1.32,0.42) -| + (1.5,0.85);
    
    \draw (ct2)+(-1,0.51) |-  (ct2);
    \draw (ct2) -| + (1,0.51);}
    \equals
    \circuit{
    init {$\ket{\psi_0}$} a;
    init {$\ket{\psi_1}$} b;
    init {$R_z(2^k\theta)\ket{+}$} c;
    init {$\ket{\psi_2}$} e; 
    init {$R_z(2^{k+1}\theta)\ket{+}$} f;
    discard d,g;

    x c;
    cnot b,c|a;
    zz (b,c);
    [name=ct1l]
    cnot d|a;
    init {} d;

    hspace {2.4cm} f;
    x f;
    cnot e,f|d;
    zz (e,f);
    cnot g|d;
    [name=ct2]
    box {$R_z(2^{k+2}\theta)$} g;
    cnot g|d;
    zz (e,f);
    cnot f|e;
    cnot e,f|d;
    hspace {5.5mm} d;
    x f;
    
    [name=ct1r]
    cnot d|a;
    discard d;
    zz (b,c);        
    cnot c|b;
    cnot b,c|a;
    x c;
    
    output {$R_z(2^k\theta)\ket{\psi_0}$} a;
    output {$R_z(2^k\theta)\ket{\psi_1}$} b;
    output {$R_z(2^k\theta)\ket{+}$} c;
    output {$R_z(2^{k+1}\theta)\ket{\psi_2}$} e;
    output {$R_z(2^{k+1}\theta)\ket{+}$} f;

    \draw (ct1l) + (-0.6,0.5) |- (ct1l);
    \draw (ct1l) -- +(0.6,0);
    \draw (ct1r) -| + (0.6,0.5);
    \draw (ct2)+(-1.78,0.65) |-  (ct2) ;
    \draw (ct2) -| + (1.78,0.65);
    }
\end{yquantgroup}
\end{adjustbox}
\caption{An example of a 2-layer in-circuit tower with `CT block' representation on the left and corresponding full circuit on the right.}
\label{fig:2layer}
\end{figure*}

Ref.\cite{sun2024} introduced a catalyst circuit gadget that can very efficiently prepare rotational resource states $R_z(\theta)\ket{+} \equiv \ket{R_z(\theta)}$. As input, the circuit takes a so-called catalyst state $\ket{R_z(\theta)}$ as well as a ``seed rotation" $\ket{R_z(2\theta)}$, and outputs $R_z(\theta)$ rotations applied to two arbitrary quantum states while also recovering the catalyst state. Therefore, the catalyst state needs to be synthesized only once and can be reused repeatedly -- the potentially high startup cost can thus averaged out over a large number of subroutine repetitions. More specifically, Ref.\cite{sun2024} introduced the following two types of \textit{catalyst towers}.

First, the in-circuit towers act directly on the computational qubits. In the multiple repetition limit (where the startup cost is averaged out), an $n$-layer in-circuit tower can apply the rotations $\{R_z(2^i \theta)\}_{i=0}^{n-1}$ at a cost of $R_T + 4n$ T states, where $R_T$ is the T-count required to synthesize a rotation gate. In contrast, the T-count for canonical gate synthesis is $n \cdot R_T$. The in-circuit towers achieve this T-count reduction by increasing the measurement depth by $2n$ compared to canonical gate synthesis. The in-circuit catalyst tower is composed of the `generalized phase catalysis circuit' introduced in \cite{gidney2019efficient}. An example of a 2-layer in-circuit tower is shown in \cref{fig:2layer}. In this case, the tower is catalysed by $\ket{R_z(2^k \theta)}$ and $\ket{R_z(2^{k+1} \theta)}$. We call the $\ket{R_z(2^{k+2} \theta)}$ gate the \textit{seed rotation} and it is applied using gate synthesis. The `corners' in the circuit are logical-AND gates defined in ref.~\cite{Gidney2018halvingcostof}, which are reproduced in \cref{appx:logicaland}. 

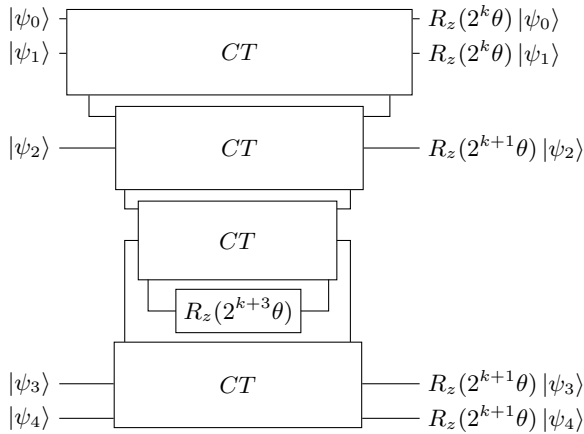
\begin{figure}
    \centering

\begin{tikzpicture}
\begin{yquant*}
qubit {$\ket{\psi_0}$} q1;
qubit {$\ket{\psi_1}$} q2;
nobit q3;
nobit a;
qubit {$\ket{\psi_2}$} b;
nobit c;
nobit d;
nobit e; 
nobit f;
nobit g;
nobit h;
qubit {$\ket{\psi_3}$} i;
qubit {$\ket{\psi_4}$} j;

box {$\qquad\qquad\qquad CT \qquad\qquad\qquad$} (q1,q2,q3);
hspace {6.5mm} b;
[name=ct0]
box {$\qquad\qquad CT \qquad\qquad$} (a,b,c);
hspace {9.5mm} e;
[name=ct1]
box {$\qquad\quad CT \qquad\quad$} (d,e,f);
hspace {14.5mm} g;
[name=ct2]
box {$R_z(2^{k+3}\theta)$} g;
hspace {6.3mm} h;
box {$\qquad\qquad CT \qquad\qquad$} (h,i,j);

output {$R_z(2^k\theta)\ket{\psi_0}$} q1;
output {$R_z(2^k\theta)\ket{\psi_1}$} q2;
output {$R_z(2^{k+1}\theta)\ket{\psi_2}$} b;
output {$R_z(2^{k+1}\theta)\ket{\psi_3}$} i;
output {$R_z(2^{k+1}\theta)\ket{\psi_4}$} j;

\draw (ct0)+(-2,0.7) |- + (-1.65,0.42) ;
\draw (ct0)+(1.65,0.42) -| + (2,0.7) ;

\draw (ct1)+(-1.5,0.67) |- + (-1.32,0.42) ;
\draw (ct1)+(1.32,0.42) -| + (1.5,0.67) ;

\draw (ct2)+(-1.2,0.42) |-  (ct2) ;
\draw (ct2) -| + (1.2,0.42) ;

\draw (ct1)+(-1.5,-1.35) |- (ct1); 
\draw (ct1) -| +(1.5,-1.35); 

\end{yquant*}
\end{tikzpicture}
    \caption{An example of a 3-layer independent catalyst tower. The catalyst states are omitted for simplicity. To produce the resource states to be used in gate teleportation, all the $\ket{\psi_i}$ states should be initialised to $\ket{+}$.}
    \label{fig:3independent}
\end{figure}

Second, the independent towers achieve reductions in both T-count and depth compared to conventional rotation gate synthesis. In this scheme, the independent towers act like higher-level magic state factories that consume T-states and produce resource states $\{\ket{R_z(2^i \theta)}\}_i$. The rotation gates in the circuit are then implemented by gate teleportation. Since the rotation angle is arbitrary, there is no efficient way to correct the `wrong' measurement outcome in the teleportation. Therefore, the success probability of this teleportation is only 0.5. The teleportation is thus repeated until success (RUS) \cite{10.5555/2685179.2685181}, but the rotation angle of the teleported state is doubled in each repetition. This scheme is probabilistic but the expected depth of the circuit is independent of $R_T$ and scales only logarithmically in $n$. An example of a 3-layer independent catalyst tower is shown in \cref{fig:3independent}. The independent towers are designed such that their yield closely caps the expected number of rotations required by the phase oracle circuits. In fact, as we will see in \cref{sec:gaussian}, the independent towers can be flexibly modified by adding or removing the CT blocks to match their output with the distribution generated by the RUS scheme.


\section{First use case: phase oracle using catalyst towers} \label{sec:poc}
As an important practical example, we estimate the cost of implementing phase oracle circuits (POC) that perform
the oracle mapping $\ket{x}\mapsto\exp(i f(x))\ket{x}$ following the low-depth construction of ref.~\cite{sun2024}.
To explicitly estimate costs, we focus on the specific application of an option pricing algorithm that we summarise in
\cref{appx:prior_work} where space-time tradeoff is highly desired, however, we note that POCs have broad applications beyond
option pricing, e.g., in grid-based quantum chemistry~\cite{hans_grid_based_2023}
and materials science~\cite{jnane_ab_initio_2024} simulations~\cite{sun2024}.

\subsection{Details of the surface code layout}

\begin{figure*}
    \centering
    \includegraphics[width=0.98\textwidth]{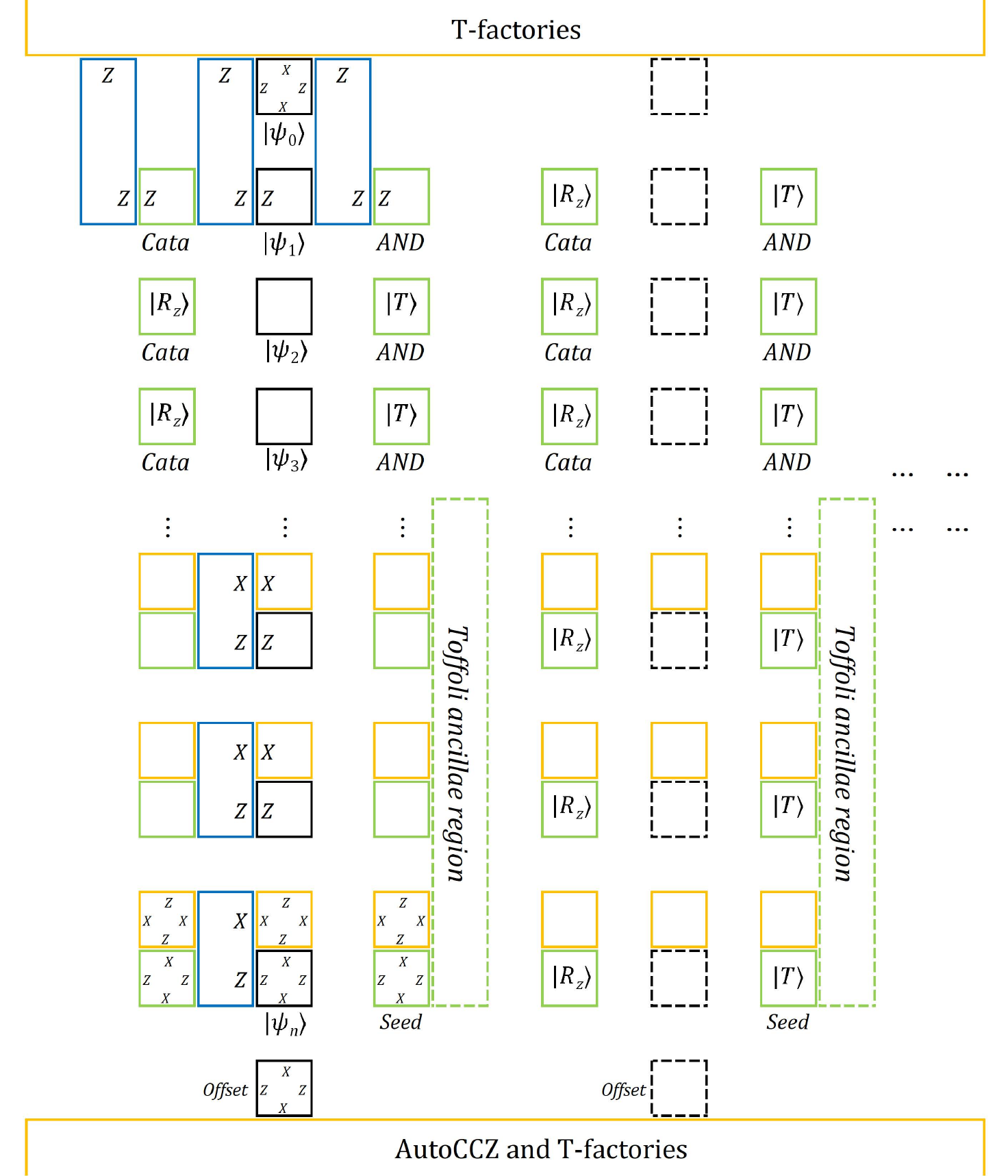}
    \caption{The layout scheme of the POC with in-circuit towers in surface code. The figure shows two towers and others are placed horizontally to the right-hand side. The black boxes are computational qubits that appear in POC, where solid line black boxes and the dashed line black boxes form two `copies' of the input quantum state to the POC separately. The green boxes are ancillary qubits for constructing the towers, where `Cata' stands for the catalyst states, while `AND' indicates that the qubit is used for performing the Logical-AND gate which starts in the T-state. Each orange square is one of three AutoCCZ magic state qubits. The green dashed line rectangles indicate the region for multi-controlled Toffoli ancillae. The $X$ and $Z$ operators of the qubit patches are indicated in the boxes. The magic state factories are at the top and bottom of the towers and the magic states are consumed via the blue rectangular ancillae.}
    \label{fig:POC}
\end{figure*}

The layout scheme of the POC is shown in \cref{fig:POC}, where the implementation uses the in-circuit towers (two towers are shown explicitly). The black boxes labelled $\ket{\psi_i}$ (and Offset) are the computational qubits in POC, where solid line black boxes and the dashed line black boxes separately form two `copies' of the input quantum state to the POC by fan-out. The green boxes are ancillary qubits for constructing the towers, where `Cata' stands for the catalyst states $\ket{R_z(2^i \theta)}$, while `AND' indicates that the qubit is used for performing the Logical-AND gate which starts in the magic state $\ket{T}$. Every orange box is one of the three AutoCCZ magic state qubits. The green dashed line rectangles indicate the region for ancillae used by multi-controlled Toffoli gates.

As we highlighted earlier, we assume a phase oracle is a subroutine that needs to be repeated many times as part of a potentially deep algorithm (e.g. an amplitude estimation algorithm) and we assume there is sufficient time to synthesise and deliver resource states before the phase oracle is called. The blue rectangles are examples of ancillae for consuming the magic states via lattice surgery. All magic states are auto-corrected using the techniques in \cite{gidney2019flexible,litinski2019game}. The $X$ and $Z$ operators of the qubit patches are indicated in the boxes. Roughly, the POC can be decomposed into five steps symmetrically: fan-out, flag computation, rotations, flag uncomputation and fan-in. We now elaborate on the implementation of POC on this layout step by step. 

Firstly, we make an improvement over the original POC here, which is also suggested in the appendix of \cite{sun2024}. For fanning out an $n$ qubit state $\ket{x_0 \ldots x_{n-1}}$ over $k$ registers, instead of using $n$ multi-target CNOT gates, we create $n$ $k$-qubit GHZ states locally. We can then distribute $k-1$ of them to locations where we wish to construct the towers (the black boxes in \cref{fig:POC}), and all these operations can be completed before the computation. When the state $\ket{x_0 \ldots x_{n-1}}$ is input during real-time computation, for every $\ket{x_i}$ where $x_i \in \{0,1\}$, we can apply a CNOT controlled on $\ket{x_i}$ and targeted at the undistributed qubit of the GHZ state to obtain the state $\ket{x_i}\frac{1}{\sqrt{2}}(\ket{x_i}\ket{0^{k-1}} + \ket{\bar{x_i}}\ket{1^{k-1}})$. Then a $Z$ measurement on the undistributed qubit (followed by a $X^{\otimes k-1}$ correction if the outcome is 1) will give the required $k$ copies. As for fan-in, $k-1$ $X$ measurement with a $Z$ correction can achieve the map $\ket{x_i}\ket{x_i^{k-1}} \mapsto \ket{x_i}\ket{0^{k-1}}$ and one can perform all the $n$ maps in parallel. Therefore, we can perform fan-in and fan-out without using the long-range multi-target CNOT gates in the original POC.

The `flag gates' are multiple-controlled Toffoli gates. An $l$-controlled Toffoli can be constructed using $l-1$ conventional Toffoli gates and at most $l-1$ ancillae \cite{he2017decompositions}. We use AutoCCZ magic states to implement the Toffoli gates and keep all the ancillae in the green dashed line rectangles. The ancillae are used to store intermediate results of computing the $l$-controlled Toffoli, which can also be used to uncompute it later (the second layer of flag gates in the POC) using $X$-basis measurements. In other words, by storing $l-1$ ancillae, we save $l-1$ CCZ magic states which require at least $3l-3$ logical patches to store, let alone the effort to produce them. 

The multi-target CNOT can be performed using the method in \cite{litinski2018lattice}, as long as a $Z$ edge of the control and an $X$ edge of each target can be accessed. 

We now turn to the in-circuit catalyst towers. In \cref{fig:POC}, each row of the tower roughly corresponds to one CT block. As can be seen from \cref{fig:2layer}, there are more operations within each CT block than operations between the blocks. Therefore, the towers operate from top to bottom and then go back to the top in a layer-by-layer fashion. The space between the patches is enough to perform the logical operations without slowing down due to routing problems, e.g. the 3 T-states in the logical-AND gate can be teleported in parallel from the T-factories above and below the towers. The $S{+}1$ in-circuit towers required by the POC are placed horizontally, the structures are almost all identical except the tower holding the input state does not need the CCZ states for the flag computation.

The layout of independent towers for POC requires a few modifications. We first note that the independent towers are magic state factories, so they are located in a designated area that surrounds the data qubits (see e.g. Figure 23 in \cite{litinski2019game}). Therefore, the independent towers do not need to be placed orderly like the in-circuit towers. They can be separated and fitted into any unfilled regions, as long as the output resource states can be delivered to the data qubits. Moreover, no CCZ states are required in the towers and thus the Toffoli ancillae region can also be removed. All the black boxes should start in $\ket{+}$ state and the `offset' patches can be removed. One key feature of the independent towers is that most of the time, two layers instead of one are operating in parallel. This can be seen from \cref{fig:3independent}, where the two CT blocks producing $R_z(2^{k+1}\theta)\ket{\psi_i}$ run in parallel. The space in the towers is sufficient to perform all logical operations without the need for additional routing of the qubits. However, it is important that the T-states are available at both the top and bottom of the towers, as in the present construction two layers are consuming T states simultaneously (the longer blue rectangles in \cref{fig:POC} now sprout from both top and bottom at the same time). 

\begin{figure*}
    \centering
    \includegraphics[width=0.8\linewidth]{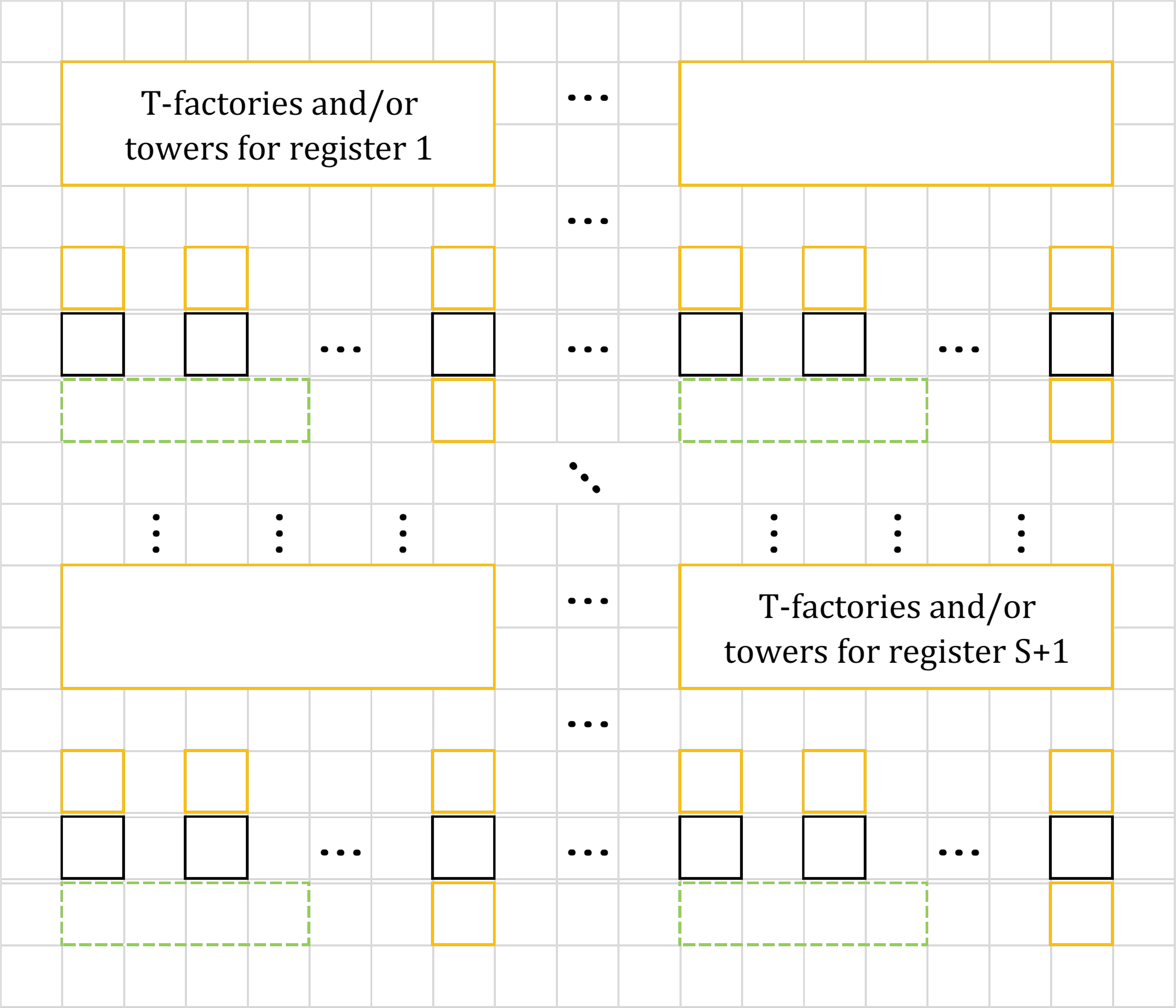}
    \caption{Layout of data qubits for gate synthesis and for the independent tower method, which supports massive parallel gate teleportation. The $S+1$ registers can be placed in a relatively dispersive way as a result of limited inter-register operations. The black boxes are data qubits and every orange square is one of the three AutoCCZ magic state qubits. The green dashed line rectangles indicate the region for multi-controlled Toffoli ancillae. The towers and/or magic state factories are placed in the orange rectangles.}
    \label{fig:data}
\end{figure*}

Now the logical data qubits are in a separate region with its own layout. Since we are teleporting a large number of resource states in parallel, it is important that the layout supports this parallelisation. A possible layout is shown in \cref{fig:data}, which ensures every data qubit is equally exposed to the distillation factories and/or independent towers. Note that since the gate synthesis method requires the same level of parallelisation to deliver T-states, it can use the same layout. As before, the black boxes are data qubits, orange squares are CCZ magic states (one out of three) and the green dashed regions are for Toffoli ancillary qubits.

\subsection{Cost Analysis}\label{sec:POCcost}

We now calculate the space overhead to implement the POC using the in-circuit towers with parameters $S$ (number of pieces of the piecewise function) and $n$ (number of qubits in the input register) on the surface code. \cref{fig:POC} visually confirms that the number of logical qubits required for $S+1$ towers
(with 1 tower needing no CCZ states) in the data block is 
\begin{equation} \label{eqn:incircuit}
    7S(2n+l+2)+7(2n+3) 
\end{equation}
where $l=\lceil \log_2{S} \rceil$. This expression includes routing qubits (the white space in the layout in \cref{fig:POC}).

For the independent tower method, the number of logical qubits is
\begin{equation}\label{eqn:independent}
    5[(S+1)(2n+2)]+4(6n+1)(S+1)+4\times4S,
\end{equation}
where the first term is the contribution of data qubits and ancillae (so for the gate synthesis method, this term is all it requires). The second term is due to the towers, i.e., we need $S+1$ towers, each of which requires $6n+1$ qubits, the factor of 4 accounts for the routing qubits and the third term is due to the 1-layer towers that deal with offset rotations.

\begin{figure*}
    \centering
    \includegraphics[width=.49\textwidth]{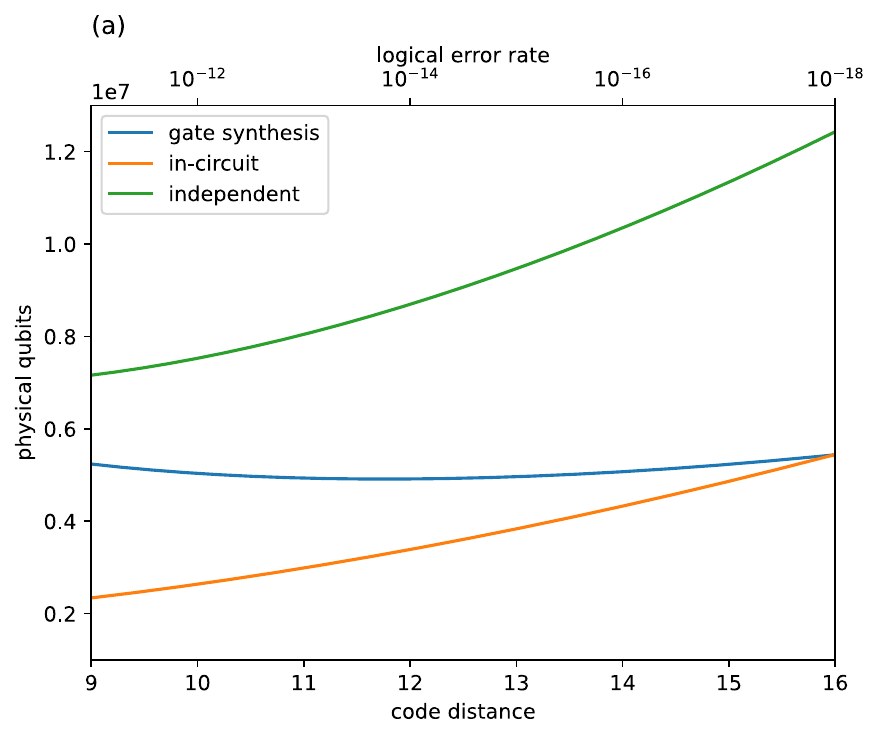}\hfill
    \includegraphics[width=.49\textwidth]{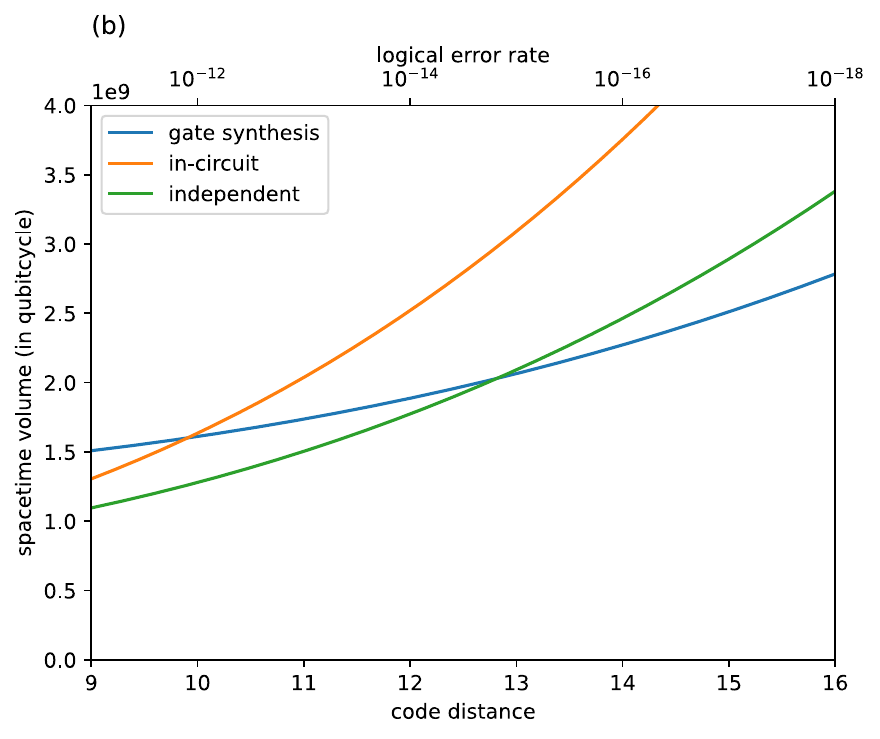}
    \caption{The (a) physical qubit count and (b) spacetime volume of the gate synthesis method, in-circuit tower method and independent tower method for the parallel piecewise phase oracle circuit applied to an option pricing algorithm (see details in \cref{appx:prior_work}) for an increasing code distance. The logical error rate of the corresponding code distance is also shown in the log scale as a secondary horizontal axis.}
    \label{fig:pocplt}
\end{figure*}

In order to estimate the space cost of distillation factories, we need to fix certain parameters explicitly. As such, we adopt the problem setting with its parameter settings for a derivative pricing algorithm from ref.~\cite{sun2024}, in which $S=36, n=15, l=6$. Furthermore, we assume that the algorithm requires a total of $\sim 10^8$ T-states (accounting for all amplitude estimation iterations) and we target a distillation error of $1\%$. We therefore require a magic state factory with output error rate below $10^{-10}$, which leads us to choose the $\text{(15-to-1)}_{11,5,5}$ protocol from \cite{litinski2019magic}, assuming a physical error rate $p\sim 10^{-4}$. A key feature of this protocol is that its code distance is independent of the logical data qubit code distance. Its spacetime volume is given by 2070 physical qubits $\times$ 30 cycles. In \cref{fig:pocplt} we compare the physical qubit count and spacetime volume of POCs using the different continuous rotation synthesis algorithms for an increasing code distance $d$ (with the corresponding logical error rate from \cref{eqn:errorrate}). See more derivation details in \cref{appx:est}. 

We deal with the code distance in the following way. Usually, an algorithm has a fixed logical qubit count and circuit depth; therefore, we can define a circuit-level probability of failure $P$, i.e., the probability of a single logical error happening in a circuit, which will require the code distance to satisfy
\begin{equation}\label{eqn:ineqn}
    \#logical\ qubit \times circuit\ depth \times p_L(d) < P,
\end{equation}
such that $d$ is the smallest odd integer that satisfies this inequality. In the present example, however, the phase oracle is a subroutine within a deeper amplitude estimation algorithm. Although the total T-count of the algorithm is fixed, the overall circuit depth depends on both the T-depth and the Clifford depth (if the Clifford gates are explicitly executed). Similarly, while the phase oracle is resource-intensive in terms of qubit count due to a width–depth trade-off, the algorithm may require even more qubits elsewhere—as is the case in the option pricing algorithm. All of these factors influence the choice of code distance. On the other hand, the logical error rate is exponentially suppressed as the code distance increases, so the impact of variation in code distance is expected to be small.


\cref{fig:pocplt}(a) indeed confirms that the in-circuit method has the lowest space cost -- but this is because all the rotations are applied in place. For the other two methods, direct gate synthesis and independent towers, a large number of states are teleported simultaneously
and we do not want them to be slowed down by the routing traffic, therefore they require extra routing space. In exchange, they are more efficient in time. The measurement depth of gate synthesis method, in-circuit tower method and independent tower method are 32, 62 and 17 time steps respectively, as calculated in \cite{sun2024}. Indeed, the independent tower approach is the shallowest (lowest time cost) since its depth scales logarithmic in $S$ and $n$.

The overall spacetime tradeoff can be inferred from \cref{fig:pocplt}(b), where we plot the total spacetime volume. Up to code distances $\sim 13$, independent towers require lower spacetime volume than conventional gate synthesis. This confirms that independent towers can be superior: not only do they provide a spacetime tradeoff -- i.e., reducing the total depth of the computation by using more qubits -- but at the same time they can even reduce the total spacetime volume. This is in contrast to common techniques that exploit spacetime tradeoffs while preserving total volume, or ones that may even incur an additional spacetime overhead~\cite{litinski2019game,webber2022impact}. The towers can still be valuable when one goes beyond the threshold code distance (13 in this case), as the independent towers allow speeding up the computation by a factor of $\sim 2$ with a modest increase in spacetime volume. However, for code distances above 16, direct gate synthesis appears to outperform the more advanced techniques in the present application example, as it requires the least space time volume and the least space.


Finally, let us highlight that, while our analysis concludes superior performance of catalyst towers for low-to-medium code distances, our analysis is specific to the present example of a derivative pricing algorithm.

\section{Second use case: catalyst towers for repeated rotations} \label{sec:gaussian}

In the previous section we considered an application that requires rotations $\bigotimes_{i=0}^{n-1} R_z(2^i a)$ applied in parallel, in which case
the in-circuit towers are advantageous in terms of T-count. Such rotations are useful, e.g., for constructing a block-encoding of the sine function~\cite{wang2023option}.

However, the independent towers can have broader applications when implementing multiple rotations of the same angle
(given the circuit is repeatedly applied) and can reduce both T-count and T-depth~\cite{kiumi2024te}.
For example, the derivative pricing algorithm detailed in \cref{appx:prior_work} initialises a number of copies of the same quantum register 
with amplitudes that follow a Gaussian distribution.
In this section we illustrate our results on this Gaussian loading approach, but another natural use case of this
could be applying Clifford hierarchy rotations via RUS as in~\cite{kiumi2024te}. 

\subsection{Details of implementation}
\begin{figure}
    \centering

\begin{tikzpicture}
\begin{yquant*}
qubit {$q1:\ket{0}$} a;
qubit {$q2:\ket{0}$} b;
qubit {$q3:\ket{0}$} c;
qubit {$q4:\ket{0}$} d;
qubit {$q5:\ket{0}$} e;

box {$R_y(\theta_{1,1})$} a;
box {$R_y(\theta_{2,1})$} b;
box {$R_y(\theta_{3,1})$} c;
box {$R_y(\theta_{4,1})$} d;
box {$R_y(\theta_{5,1})$} e;
cnot b|a;
cnot d|c;
cnot c|b;
cnot e|d;

text {$\dots$} a,b,c,d,e;

box {$R_y(\theta_{1,7})$} a;
box {$R_y(\theta_{2,7})$} b;
box {$R_y(\theta_{3,7})$} c;
box {$R_y(\theta_{4,7})$} d;
box {$R_y(\theta_{5,7})$} e;
cnot b|a;
cnot d|c;
cnot c|b;
cnot e|d;

\end{yquant*}
\end{tikzpicture}
\caption{A variational $R_y$-CNOT circuit is trained classically in~\cite{PhysRevA.98.022322} to prepare a quantum state whose amplitudes follow a
 normal distribution. $60$ copies of this circuit are used in the derivative pricing algorithms (detailed in \cref{appx:prior_work}) to initialise
 60 registers in a multivariate normal distribution. We use the identity $R_y(\theta)=SHR_z(\theta)HS^{\dagger}$ and estimate the cost of preparing the continuous rotations.}
\label{fig:gaussian}
\end{figure}
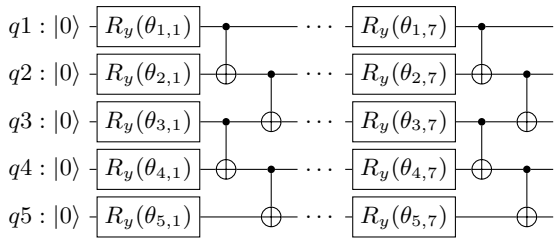

The option pricing algorithm detailed in \cref{appx:prior_work} is initialised by loading a standard normal distribution onto a quantum register, i.e., 
a classically trained $R_y$-CNOT ansatz~\cite{PhysRevA.98.022322} is used (see \cref{fig:gaussian}) to prepare a quantum state whose amplitudes follow
a normal distribution. Each variational circuit uses 5-qubits and has 7 layers with 35 parametrised $R_y$ rotations --
the algorithm then prepares 60 quantum registers in parallel by executing 
coherent copies of this circuit, which in total requires 2100 rotations. The independent towers can be used to
prepare the rotation resource states which are then teleported via RUS.
The independent towers are also modified (compared to \cref{fig:3independent}) to match the geometric distribution of RUS and a family of towers of various layers is needed for each different variational angle $\theta_{i,j}$. For example, a 3-layer tower, in this case, has an extra CT block attached to the $\ket{\psi_4}$ qubit line in \cref{fig:3independent}, producing 4 $\ket{R_z(2^{k}\theta)}$ and 2 $\ket{R_z(2^{k+1}\theta)}$ states. In general, this construction is suitable for most circuits that implement multiple identical rotations.

We fix the accuracy of single rotation to be $\epsilon=2\times10^{-6}$ and require $r=200$ repetitions of the state preparation oracle as part of amplitude estimation.
We estimate that canonical gate synthesis requires 53024 T-states per repetition and has a measurement depth of 177.
We then construct two variants of independent towers. First, the \emph{control} approach produces minimal excess resource
states while requiring several towers of different heights. Second, the \emph{excess} approach minimizes the number of towers required
but consumes more T-states and produces many redundant resource states. 
These two variants of independent tower constructions then require an average of 28709 (control scheme) and 45750 (excess scheme) T-states
per repetition and both have an expected measurement depth of 39 (see \cref{appx:depth} for details).
As in our previous application example, 
we are again in a situation that requires a number of parallel state teleportations, and therefore a layout similar to \cref{fig:data} can be used. 

\subsection{Cost Analysis}\label{sec:gausscost}

We first note that while the rotations for Gaussian state preparation using independent towers can be performed in 39 measurement steps using RUS,
the resource-state preparation stage, i.e., running the catalyst tower circuits, requires more measurement steps 
(the measurement depth of an $L$-layer independent tower is $R_T + 2L$). 
For this reason we need to assume a produce-and-store strategy for the rotational resource states whereby
the towers are running throughout the course of the algorithm and the resulting resource states are stored to be used later by 
high resource consuming subroutines. 

\begin{figure}
    \centering
    \includegraphics[width=\linewidth]{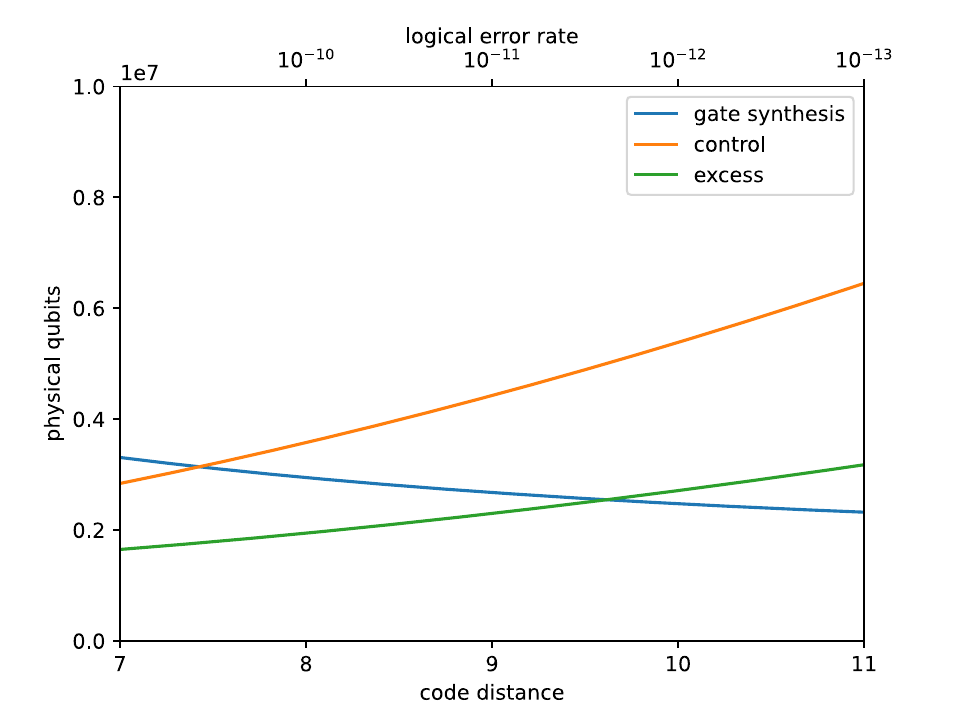}
    \caption{Physical qubit count of conventional gate synthesis, independent towers with the control scheme and
    	independent towers with the excess scheme for Gaussian state preparation for an increasing code distance.
    	The logical error rate of the corresponding code distance is also shown in the log scale as a secondary horizontal axis.}
    \label{fig:gaussplt}
\end{figure}

We assume every iteration of amplitude estimation takes $\sim10^3 d$ code cycles and we use the  $\text{(15-to-1)}_{11,5,5}$ magic state factory 
as in the previous section. For a total of 300 abstract algorithmic qubits (60 copies of 5-qubit circuits) we require 1200 logical data qubits to account for routing space between the algorithmic qubits, i.e., we assume a ratio of 1:3 of algorithmic to routing qubits. Similarly, while an $L$-layer catalyst tower requires $6L-2$ abstract algorithmic qubits, when we include routing space, these towers have a total footprint of approximately $4(6L-2)$ logical qubits. Furthermore, in the present example, the tower approach requires an additional 4200 logical qubits for temporarily storing resource states, i.e., 2100 rotations multiplied by a factor of 2 to account for the success rate (0.5) of the RUS approach. The physical qubit count as a function of code distance is shown in \cref{fig:gaussplt}.

For code distances below 10, the excess scheme uses fewer physical qubits while being $\sim4.5$ times faster. Consistent with our findings in the previous section, we conclude that both the control and excess tower schemes offer greater efficiency than conventional gate synthesis at small code distances. Given that the code distance of distillation factories is independent of that of the logical data qubits, this implies that direct T-state production may be more cost-effective than catalytic approaches at large code distances.

\section{Discussion} \label{sec:discussions}
In this section, we discuss several factors that may influence our conclusions, as well as the potential for generalising the present approach.\\

\noindent
\textbf{Sensitivity to parameter settings ---} 
Let us first emphasize that, in the present study, we made several parameter choices motivated by a specific practical application. For example, we fixed the total T-count of the algorithm, which in turn allowed us to select a particular T-factory. Additionally, we employed a fallback protocol for gate synthesis—whose T-count is one-third that of the optimal, deterministic, and ancilla-free protocol \cite{kliuchnikov2023shorter}—and we assumed a routing factor of 4 to support a high degree of parallelism. We evaluated the space and time complexity of the algorithm under different rotation gate implementation strategies; however, we expect that our conclusions may vary under a different set of hyperparameters or in an alternative problem setting.
\\

\noindent
\textbf{Distillation cost vs Data qubit cost ---} 
A range of algorithmic primitives and protocols have been developed that formally reduce the T-gate count of continuous rotations, such as Hamming weight phasing~\cite{kivlichan2020improved} and catalyst towers \cite{sun2024}, where the latter was the focus of the present study. However, as we demonstrate in the present work, even though these techniques formally reduce T counts at a high-level circuit description, as soon as overheads due to hardware constraints and quantum error correction 
are taken into account, the superiority of these techniques may diminish. Let us illustrate this with the application example considered in this work, where magic state production is not the time-limiting factor. In this case, any reduction in the number of required magic states is equivalent to a decrease the number of distillation factories, which in turn directly reduces the overall physical qubit requirements. However, the implementation of T-state-saving gadgets, such as Hamming weight phasing or catalytic circuits, requires additional logical ancilla qubits. If the distillation factories operate at a fixed code distance that does not scale with the overall system size -- as assumed in this study -- then, at large code distances, the savings in physical qubits from reducing the number of distillation factories may be offset by the $O(d^2)$ physical qubits required to implement the ancilla logical qubits, as observed in our 'control vs excess' schemes discussed in \cref{sec:gausscost}. On the other hand, additional factors may significantly impact the space complexity of magic state preparation. For instance, deeper circuits typically require larger code distances, and assuming a constant density of continuous rotations, an increase in code distance implies a higher number of required T gates -- thereby necessitating a more costly distillation factory. Consequently, the code distance of the distillation factory can be expected to implicitly depend on that of the data qubits.\\

\noindent
\textbf{Parallelisation and routing space ---} 
This work is focused on an application example whereby the primary aim is to reduce the time complexity by introducing a minimal space overhead. Such spacetime tradeoff is expected to be desired for a broad range of applications where the time to answer is required to be minimal, e.g., in applications where only polynomial speedups over classical algorithms is possible. However, we observe that our catalyst towers in \cref{fig:pocplt}b not only reduce the time complexity, but
also reduce the overall spacetime volume of the computation, suggesting that for low to medium code distances this is a worthwhile approach. Achieving this superior speed also requires us to dedicate generous routing space in order for the algorithm not to be slowed down by routing traffic. Therefore, we used a 1:3 computational to routing qubit ratio that ensures every edge of the data qubit patch is exposed, which is sufficient for implementing logical operations within the catalyst circuits. When designing the layout of the data qubits in \cref{fig:data}, we rely on the fact that the data qubits (black squares) in POC exhibit limited interaction between registers. A similar property holds for Gaussian state preparation, where the 60 circuit copies shown in \cref{fig:gaussian} operate independently with no interaction. Without this characteristic, our layout would be inefficient for long-range inter-register operations.\\

\noindent \textbf{Generalisations and early fault tolerance ---}
In \cref{sec:gaussian}, we designed a layout that implements multiple copies of a variational circuit for preparing Gaussian quantum states, as shown in \cref{fig:gaussian}. This approach can be readily extended to a broad class of VQE ansatz circuits, i.e., variational circuits composed of gate blocks with identical rotation angles, such as the variational Hamiltonian ansatz or QAOA~\cite{cerezo2021variational, Boyd_2022}. Furthermore, our approach can be straightforwardly generalized to broader classes of ansaetze featuring redundant rotation angle settings. In the limiting case, our results are directly applicable to quantum dynamics simulations based on randomized product formulas, such as qDRIFT~\cite{campbell2019random} or TE-PAI~\cite{kiumi2024te}, where the same rotation angle is applied across all quantum gates in the circuit. Notably, Ref.~\cite{kiumi2024te} constructed optimized catalyst circuits in this setting and demonstrated their practical advantages. As such, our approach may serve as an enabling technique for several of the most prominent early fault-tolerant applications, including quantum dynamics simulation, robust phase estimation or spectroscopy~\cite{gunther2025phase, PRXQuantum.6.010352}, variational quantum algorithms~\cite{cerezo2021variational, Boyd_2022}, and beyond. In fact, early fault tolerance represents a particularly promising domain for our methods, for two key reasons. First, early fault-tolerant devices are typically constrained to low or intermediate code distances, where our techniques demonstrate a clear advantage. Second, these applications often benefit from spacetime trade-offs aimed at reducing the runtime of individual circuit executions, since early fault-tolerant algorithms generally require many repeated executions.

\section{Conclusion} \label{sec:conclusion}

In this work, we address the question: ``What is the most efficient approach for implementing rotations in fault-tolerant settings?" While several gadgets, such as Hamming weight phasing~\cite{gidney2019efficient} and catalyst towers~\cite{sun2024,wang2023option,kiumi2024te}, have been developed to reduce the total T-count of rotations, we argue that the ultimate cost function one needs to optimise against should be either the total runtime of the algorithm (time complexity) or the number of physical qubits required (space complexity). When all overheads associated with hardware constraints and fault tolerance are taken into account, gadgets optimized solely for minimal T-count may not, in fact, yield the most efficient implementation.

We develop explicit implementations of catalyst towers that minimise the space and time complexity of rotations in a surface code architecture. We carefully arrange logical qubits in our layout and dedicate sufficient routing space to enable a high degree of parallelization. We then compare the cost of different rotation gadgets in two specific application examples, both of which are subroutines in usual option pricing applications~\cite{chakrabarti2021threshold, grinko2021iterative, stamatopoulos2024derivative}. First, we consider a phase oracle, which is a ubiquitous algorithmic primitive in a broad range of applications, and focus on a low-depth, parallel piecewise oracle as in \cite{sun2024}. Second, we consider the implementation of a variational quantum circuit that is used to prepare a number of copies of a Gaussian state.

In the specific application examples considered in this work, we find that catalyst towers are not only faster but can also reduce the total spacetime volume required for implementing rotations at low and intermediate surface code distances. In contrast, conventional Clifford$+$T gate synthesis appears to be the most efficient approach at large code distances. However, we emphasize that these conclusions are sensitive to both the specific application and the hyperparameter choices made. Nevertheless, as discussed above, our catalyst tower constructions are broadly applicable to a wide range of use cases, including quantum dynamics simulation~\cite{campbell2019random,kiumi2024te}, spectral estimation via robust phase estimation or spectroscopy~\cite{gunther2025phase, PRXQuantum.6.010352}, variational quantum algorithms~\cite{cerezo2021variational, Boyd_2022}, and beyond. As such, our techniques may serve as enabling tools for many early fault-tolerant applications, where low to intermediate code distances are typically assumed and spacetime tradeoffs are often desirable.

While optimizing a quantum circuit using lattice surgery is known to be NP-hard~\cite{herr2017optimization}, the layouts presented in this work were developed manually, which may have resulted in some loss of efficiency, e.g., in favor of clarity or readability. As an illustration, in \cref{fig:POC}, the patches are arranged in a tower-like structure, which may introduce redundant routing patches, suggesting that further optimization is possible. Furthermore, adopting the magic state cultivation approach of ref.~\cite{gidney2024magic} would likely reduce the spacial and temporal cost of magic state production. On one hand this would limit the effectiveness of towers to even lower code distances, however, on the other hand, the logic of laying out circuits on surface code might change significantly, as the usual ``data block surrounded by distillation block'' approach may be replaced by e.g., in-place magic state production.

We also highlight that future research should focus on the development of quantum compilers capable of incorporating different resource optimization priorities -- such as spacetime tradeoffs -- and of efficiently mapping abstract quantum circuits onto surface code architectures. Additionally, evaluating the performance of catalyst towers in the context of alternative quantum error-correcting codes beyond the surface code represents another promising direction for further investigation.

\section*{Acknowledgements}

The authors thank Simon Benjamin and Benjamin Pring for helpful technical discussions. BK thanks UKRI for the Future Leaders Fellowship Theory to Enable Practical Quantum Advantage (MR/Y015843/1). The authors also acknowledge funding from the EPSRC projects Robust and Reliable Quantum Computing (RoaRQ, EP/W032635/1) and Software Enabling Early Quantum Advantage (SEEQA, EP/Y004655/1).

\appendix

\section{The logical-AND gate} \label{appx:logicaland}
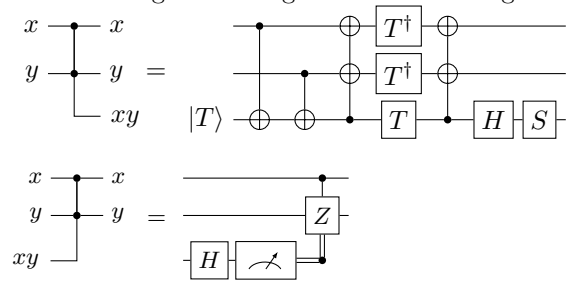
\begin{figure}[h]   
\hspace*{-4.2cm}
\begin{tikzpicture}
\begin{yquantgroup}
\registers{
    qubit {} a;
    qubit {} b;
    qubit {} c;
}
\circuit{
    init {$x$} a;
    init {$y$} b;
    discard c;
    
    [name=d1]
    zz (a,b);
    
    output {$x$} a;
    output {$y$} b;

    \draw (d1) |- + (-0.35,-1.1); 
    \node at ($(d1) + (-0.68,-1.1)$) {$xy$};
}
\equals
\circuit{

    h c;
    measure c;
    box {$Z$} b|a,c;
    discard c;    
}

\end{yquantgroup}
\end{tikzpicture}
\caption{The circuit diagram for the `corner' notation, reproducing Figure 3 of \cite{Gidney2018halvingcostof}. }
\label{fig:AND}
\end{figure}

The corner notation for computation and uncomputation of the Logical-AND gate is shown in \cref{fig:AND}.

\begin{tikzpicture}
\begin{yquantgroup}
\registers{
    qubit {} a;
    qubit {} b;
    qubit {} c;
}
\circuit{
    init {$x$} a;
    init {$y$} b;
    discard c;
    
    [name=d1]
    zz (a,b);
    
    output {$x$} a;
    output {$y$} b;

    \draw (d1) |- + (0.35,-1.2); 
    \node at ($(d1) + (0.68,-1.23)$) {$xy$};
}
\equals
\circuit{
    init {$\ket{T}$} c;

    cnot c|a;
    cnot c|b;
    cnot a,b|c;
    box {$T^\dag$} a;
    box {$T^\dag$} b;
    hspace {0.5mm} c;
    box {$T$} c;
    cnot a,b|c;
    h c;
    box {$S$} c;
}

\end{yquantgroup}
\end{tikzpicture}

\section{Option pricing algorithm using low-depth phase oracles} \label{appx:prior_work}

In this section, we introduce some details of the option pricing algorithm, which is the main focus of this work.

The approach we consider builds upon the work of Chakrabarti et al.\cite{chakrabarti2021threshold}, who proposed a quantum algorithm for option pricing based on amplitude estimation. The algorithm begins by preparing a set of quantum registers in states whose amplitudes approximate a standard normal distribution, achieved through variationally trained state preparation circuits. An affine transformation is then applied to these distributions via quantum arithmetic circuits to represent possible future `paths' of assets. This circuit is subsequently repeated as part of Iterative Quantum Amplitude Estimation (IQAE) protocol\cite{grinko2021iterative} to estimate the expected payoff. The primary bottleneck of this approach lies in the oracle proposed in ref.~\cite{chakrabarti2021threshold}, which computes the payoff function arithmetically and is computationally expensive. To address this, Stamatopoulos et al.\cite{stamatopoulos2024derivative} introduced an improved oracle constructed using quantum signal processing (QSP). However, conventional QSP is a strictly serialized algorithm, leading to quantum runtimes that are not expected to be competitive with those of classical simulators.

To reduce the runtime of the oracle circuits, Sun et al.~\cite{sun2024} recently introduced a low-depth implementation of general phase oracles using a parallel piecewise circuit. This speedup is achieved by fanning out a small register into multiple ``copies" and applying parts of the oracle in parallel to these copies. This approach represents a spacetime tradeoff, which, in the limiting case, enables a fully parallelized oracle with a rotation depth of just 1.

Furthermore, Sun et al.~\cite{sun2024} introduced two types of catalyst towers that can outperform conventional rotation synthesis in terms of both T-count and T-depth, particularly when the oracle circuits are repeatedly applied, such as in subroutines for amplitude estimation or in first-quantized Hamiltonian simulation algorithms.

\section{Estimating the cost of POC} \label{appx:est}

For a quantum subroutine that consumes $N_T$ T-states in $t\cdot d$ code cycles ($d$ is the code distance), if the distillation cost is $N_{fac}$ physical qubits and $t_{fac}$ code cycles for every factory, then assuming a constant rate of T-states consumption, the number of physical qubits required to run distillation factories is 
\begin{equation}\label{eqn:T_phy}
    \frac{N_T N_{fac} t_{fac}}{t} d^{-1}
\end{equation}

Denoting the number of logical qubits as $N_L$, which is discussed in \cref{sec:POCcost}, the total physical qubit count is roughly 
\begin{equation}\label{eqn:total_phy}
    \frac{N_T N_{fac} t_{fac}}{t}d^{-1}+N_L \cdot 2d^2
\end{equation}
where we used that each logical qubit requires $\sim 2d^2$ physical qubits.

The spacetime volume in qubitcycle is roughly given by
\begin{equation}\label{eqn:vol}
    N_T N_{fac} t_{fac}+ 2N_L t d^3
\end{equation}

Note that we have ignored some minor space costs in our estimate, e.g. the space for AutoCCZ factories is small compared to the T-factories.

\section{Expected T-count and depth of Gaussian state preparation} \label{appx:depth}

To implement $N$ identical $R_z(\theta)$ rotations via RUS, it is expected that we need $N$ $\ket{R_z(\theta)}$ states, $\lceil N/2\rceil$ $\ket{R_z(2\theta)}$ states, $\lceil N/4\rceil$ $\ket{R_z(4\theta)}$ states, all the way up to 1 $\ket{R_z(2^{\lceil \log_2 N \rceil}\theta)}$ state (we truncate the `higher order' rotations). In our case, we need to implement 60 copies of 35 different rotations. An independent tower of $L$ layers in this case (as described in \cref{sec:gaussian}), can produce 4 $\ket{R_z(\theta)}$ states, and 2 $\ket{R_z(2\theta)}$ states, 2 $\ket{R_z(4\theta)}$ states, ..., and 2 $\ket{R_z(2^{L-2}\theta)}$ states. Therefore for $N=60$, the required number of rotations can be provided by 7 3-layer independent towers (either construct 7 towers or run 1 tower 7 times), 4 4-layer towers, 2 5-layer towers, 1 6-layer tower and 1 8-layer tower. Since the towers can only produce an even number of states, there will be 2 excess states. The excess states can either be discarded or in practice, one might want some excess states to account for the statistical fluctuations. In our case, the T-count of an $L$-layer tower is 
\begin{equation}\label{eqn:T_tower}
    2L R_T + 4 (2L-1) + (r-1) [R_T + 4 (2L-1)]
\end{equation}
where $r$ is the rounds of repetition.

Using the fallback protocol in \cite{kliuchnikov2023shorter}, the T-count for gate synthesis is 
\begin{equation}\label{eqn:T_fallback}
    R_T \approx 1.03\log_2{(1/\epsilon)+5.75}
\end{equation}
where we set $\epsilon=2\times10^{-6}$. Therefore, the expected T-count per repetition is 28709. The above is a `control' scheme which has minimum excess resource states so the T-count is also minimized. Alternatively, an `excess' scheme can run an 8-layer tower 15 times, spending 45750 T-states per repetition but saving the logical qubits for constructing the towers.

The problem of finding the expected depth can be described as this: 5 coins are tossed at the same time, the heads are kept and the tails are re-tossed till they all become heads. Let the random variable $X$ be the depth of this RUS procedure (its pdf is explicitly derived in \cite{sun2024}), a single copy of the $R_y$-CNOT circuit repeats this procedure 7 times. Define the random variable $Y=X_1+...+X_7$ where all $X_i$ are identical, we are interested in the expectation value of $Z = max\{Y_1, ..., Y_{60}\}$, where all $Y_i$ are identical. The $max$ function is due the fact that the algorithm cannot proceed until all 60 copies of $R_y$-CNOT circuits complete all 7 layers of rotations. We use numerical simulation with 10000 samples and find $\mathbb{E}[Z]$ is approximately 39 (this code is available from the authors on request). 


\bibliography{bib}


\end{document}